\documentclass[%
 aip,
 amsmath,amssymb,
 reprint,%
]{revtex4-1}

\usepackage{graphicx}
\usepackage{dcolumn}
\usepackage{bm}

\usepackage[utf8]{inputenc}
\usepackage[T1]{fontenc}
\usepackage{mathptmx}

\begin{document}

\preprint{AIP/123-QED}

\title[A model system of coupled microcavities]{A simple model system to study coupled photonic crystal microcavities}
\title[]{A simple model system to study coupled photonic crystal microcavities}

\author{A. Perrier}
\author{Y. Guilloit}%
\author{E. Le Cren}%
\affiliation{Université de Rennes 1, IUT de Lannion, rue Edouard Branly, 22300 Lannion, France}
\author{Y. Dumeige}
\email{yannick.dumeige@univ-rennes1.fr}
\affiliation{Université de Rennes 1, IUT de Lannion, rue Edouard Branly, 22300 Lannion, France}
\affiliation{Université de Rennes 1, Institut FOTON, 6 rue de Kerampont, 22300 Lannion, France}
 
\date{\today}

\begin{abstract}
In this paper, we designed and experimentally studied several systems of standard coaxial cables with different impedances which mimic the operation of so called photonic structures like coupled photonic crystal microcavities. Using elementary cells of half-meter long coaxial cables we got resonances around $100~\mathrm{MHz}$, a range of frequencies that can be easily studied with a standard teaching laboratory apparatus. Resonant mode frequency splitting has been obtained in the case of double and triple coupled cavities. A good agreement between experimental results and transfer matrix model has been observed. The aim here is to demonstrate that standard coaxial cable system is a very cheap way and an easy to implement structure to explain to undergraduate students complex phenomena that usually occur in the optical domain.
\end{abstract}

\maketitle

\section{\label{intro}Introduction}

Optical micro-resonators are of great interest for fundamental studies in optics and for applications in photonics such as integration of optical sources, optical filtering or bio an chemical sensing \cite{Vahala03, Chiasera10}. One very popular method to integrate optical micro-resonators is to create a defect in a periodic photonic crystals \cite{Foresi97,Centeno00,Sauvan05,Kuramochi06,Combrie08}. As it is the case of atomic crystals, the periodicity breaking creates resonant modes localized in the defect with a resonance frequency lying within the photonic bandgap\cite{Yablonovitch91,Fan95}.
The coupling of optical micro-resonators gives additional degrees of freedom and enables the design of complex photonic structures, with optimized optical characteristics\cite{Karle04,Maes06,Xu06,Boriskina06}. As the states of photons confined in an optical microcavity are similar to confined electron states in atoms, optical micro-resonators are often referred to photonic atoms. In this physical picture, coupled resonators will support hybridized states and can be compared to photonic molecules\cite{Bayer98,Haddadi14,Yang17,Smith20}. From a fundamental point of view these objects are still the subject of intense research efforts in quantum photonics, nonlinear optics or laser physics\cite{Hamel15,Rodriguez16,Marconi18,Ceppe19}. The interaction of the resonant modes of two cavities leads to interesting phenomena such as frequency splitting or induced transparency and can be used for dispersion management\cite{Xue15} or optical storage\cite{Smith04,Maleki04,Dumeige08b,Peng14}. The symmetric frequency splitting obtained in a photonic molecule composed of three coupled resonators made of whispering gallery mode resonators or photonic crystals microcavities has been used to reach the phase matching condition in the four wave mixing process\cite{Azzini13,Zeng15,Armaroli15}. By increasing the number of coupled resonators it is even possible to obtain resonant waveguides\cite{Yariv99} or delay lines with applications in optical signal buffering\cite{Notomi08,Dumeige09}.\\
\indent From another point of view, coaxial cable structures have been used to emulate one-dimensional photonic crystals, Bragg mirrors or quasi-periodic photonic structures in the radio-frequency domain\cite{Hache02,Hache04,Boudouti07,Boudouti07b}. Doing a periodic system consisted of two sets of meter long coaxial cables, it is possible to observe a Bragg diffraction due to reflections at impedance transitions\cite{Sanchez03}. Adding an extra cable in the middle of the system, a Fabry-Perot type resonance appears in the center of the stop-band; this effect has already been demonstrated in the range $10-50~\mathrm{MHz}$ using coaxial cables\cite{Schneider01,Sanchez03}. From an educational point of view, the great asset of this approach is that students can build their own mirrors or cavities without the need of complex technological facilities. This is not possible in the optical domain since in this case the involved wavelengths are very short. In this paper we propose to extend and generalize this radio-frequency (RF) analogy to coupled resonator structures made of one-dimensional photonic crystal defect cavities and show that usual laboratory equipment can be used to teach the basics of nanophotonic circuitry at the undergraduate level.\\
\indent The paper is organized as follows: in section \ref{theorie} we introduce the transfer matrix method\cite{Bendickson96} (TMM) which will be used all along the paper to model our structures. Then, we detail the physics underlying the coupling of optical cavities focusing on the case of two and three defects. Section \ref{experi} is devoted to the experimental demonstration of the coupling of model photonic crystal cavities consisted of coaxial cables with two different impedances of $50~\Omega$ and $75~\Omega$.

\section{\label{theorie}Theory}\label{theorie}

\subsection{Periodic structure}

\begin{figure}
\includegraphics[width=6.5cm]{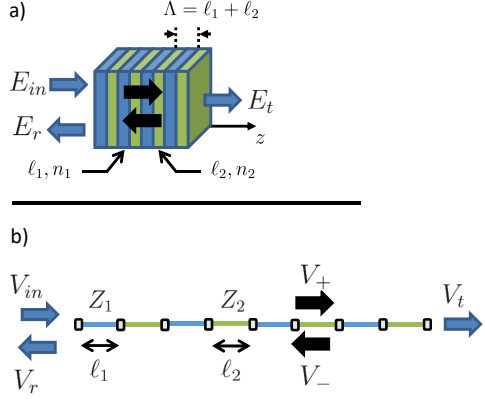}
\caption{\label{Fig1} a) Finite one-dimensional Bragg mirror or photonic crystal of period $\Lambda$. b) Analog of the photonic crystal made of two coaxial cables of impedance $Z_1$ and $Z_2$ and length $\ell_1$ and $\ell_2$. $z$ is the space coordinate, $z=0$ at the input of the strucutre.}
\end{figure}

\begin{figure}
\includegraphics[width=8cm]{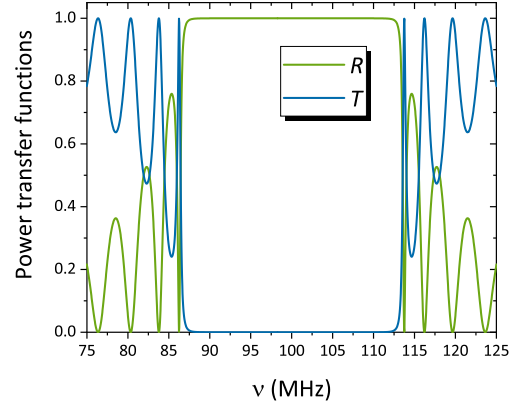}
\caption{\label{Fig2} Power transmission ($T$) and reflection ($R$) for a periodic structure made of $N=20$ cells with $Z_1=50~\Omega$, $Z_2=75~\Omega$, $v_{\phi 1}=v_{\phi 2}=\frac{2c}{3}$ and $\ell_1=\ell_2=50~\mathrm{cm}$.}
\end{figure}

Figure. \ref{Fig1}.a) shows a finite one-dimensional photonic crystal or Bragg mirror made of $N$ identical cells constituted by two dielectric material layers of refractive indices $n_1$ and $n_2$ and thicknesses $\ell_1$ and $\ell_2$. $E_{in}$, $E_r$ and $E_t$ are respectively the input, reflected and transmitted optical fields. To obtain a maximal reflection at frequency $\nu_0$ the phase accumulated by the wave after propagation within a unit cell have to be equal to $\pi$ and thus the following condition must be verified\cite{Bendickson96}:
\begin{equation}
n_1\ell_1+n_2\ell_2=\frac{\lambda_0}{2},
\end{equation}
where $\lambda_0$ is the Bragg wavelength.
The radio-frequency (RF) analog of this Bragg mirror can be obtained by a periodic system whose elementary cell consists of two coaxial cables of different length and impedance. In the RF domain it is more convenient to use voltages and we define here $V_{in}$, $V_r$ and $V_t$ as the input, reflected and transmitted voltages respectively. By introducing the phase velocity $v_{\phi 1}$ and $v_{\phi 2}$ of the cable of impedance $Z_1$ and $Z_2$, it is possible to write the Bragg condition as
\begin{equation}\label{Bragg_cond}
\frac{\ell_1}{v_{\phi 1}}+\frac{\ell_2}{v_{\phi 2}}=\frac{1}{2\nu_0},
\end{equation}
where $\nu_0$ is the resonant frequency. The propagation along the structure can be modeled using the TMM\cite{Hashizume95,Bendickson96,Sanchez03}. The reflection and transmission at each interface between two media of impedance $Z_i$ and $Z_j$ is obtained via the matrix $\mathrm{\mathbf{M}}_{i,j}$ defined by 
\begin{equation}
\mathrm{\mathbf{M}}_{i,j} =\frac{1}{t_{j,i}} 
\begin{pmatrix}
1 & r_{j,i} \\
r_{j,i} & 1 \\
\end{pmatrix}
\end{equation}
where $r_{j,i}=\frac{Z_i-Z_j}{Z_j+Z_i}$ and $t_{j,i}=\frac{2Z_i}{Z_j+Z_i}$. At angular frequency $\omega=2\pi\nu$, the propagation in a layer of thickness $\ell_i$ and phase velocity $v_{\phi i}$ is taken into account thanks to matrix $\mathrm{\mathbf{Q}}_{i}$ which reads
\begin{equation}
\mathrm{\mathbf{Q}}_{i} = 
\begin{pmatrix}
\exp{\left(j\omega\ell_i/v_{\phi i}-\kappa_i\ell_i\right)} & 0 \\
0 & \exp{\left(-j\omega\ell_i/v_{\phi i}+\kappa_i\ell_i\right)} \\
\end{pmatrix}
\end{equation}
where $\kappa_i$ is the attenuation coefficient of the medium of impedance $Z_i$. We define the period of the photonic crystal $\Lambda=\ell_1+\ell_2$; $V_+(z)$ and $V_-(z)$ are the signals propagating respectively in the forward and backward directions (see Fig. \ref{Fig1}). Assuming that $m\in\mathbb{N}$, the matrix $\mathrm{\mathbf{M}}$ associated to the unit cell defined by
\begin{equation}
\begin{pmatrix}
V_+([m+1]\Lambda) \\
V_-([m+1]\Lambda) \\
\end{pmatrix}=\mathbf{M}
\begin{pmatrix}
V_+(m\Lambda) \\
V_-(m\Lambda) \\
\end{pmatrix},
\end{equation}
is thus given by
\begin{equation}
\mathbf{M}=\mathrm{\mathbf{M}}_{2,1}\mathrm{\mathbf{Q}}_{2}\mathrm{\mathbf{M}}_{1,2}\mathrm{\mathbf{Q}}_{1}.
\end{equation}
We consider now that input and output media have both an impedance $Z_1$, thus the matrix $\mathrm{\mathbf{M}}_{\mathrm{Bragg}}$ of a periodic structure made of $N$ elementary cells is given by
\begin{equation}
\mathrm{\mathbf{M}}_{\mathrm{Bragg}}=\mathrm{\mathbf{M}}^N.
\end{equation}
In the linear regime, the voltage \textit{amplitude} transmission $t_0$ and reflection $r_0$ can be deduced by using the following relation 
\begin{equation}
\begin{pmatrix}
t_0 \\
0 \\
\end{pmatrix}=\mathbf{M}_{\mathrm{Bragg}}
\begin{pmatrix}
1 \\
r_0 \\
\end{pmatrix}.
\end{equation}
The power transmission $T$ and reflection $R$ coefficients are thus obtained by $T=\left|t_0\right|^2$ and $R=\left|r_0\right|^2$. Figure \ref{Fig2} gives an example of the transmission and reflection spectra of a 20-cells periodic structure made of coaxial cables with the following parameters: $Z_1=50~\Omega$, $Z_2=75~\Omega$, $v_{\phi 1}=v_{\phi 2}=\frac{2c}{3}$ and $\ell_1=\ell_2=50~\mathrm{cm}$.
We consider in this section a loss-less material ($\kappa_i=0$) for a sake of clarity. The propagation is forbidden within a spectral range of $27~\mathrm{MHz}$ centered at $\nu_0=100~\mathrm{MHz}$, see Eq. (\ref{Bragg_cond}), this effects manifests itself by a high reflection and a low transmission.

\subsection{Defect modes in one-dimensional photonic crystal structures}

\begin{figure}
\includegraphics[width=9cm]{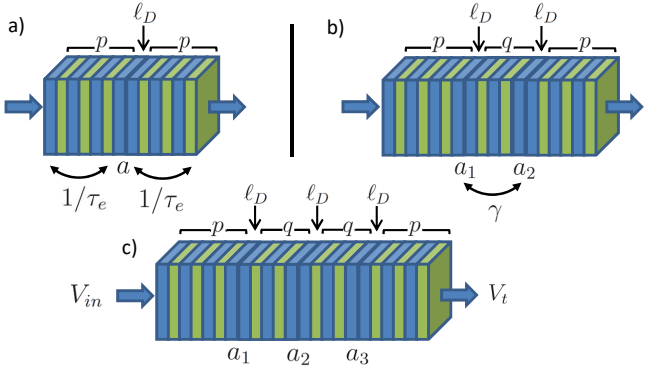}
\caption{\label{Fig3} Defect resonant structures obtained in a one-dimensional photonic crystal: a) one-dimensional photonic crystal cavity, b) two coupled cavities and c) three coupled cavities. $p$ is the number of elementary cells of the confinment barriers, the defects have an impedance $Z_D=Z_1$ and a length $\ell_D=\ell_1$, $q$ is the number of elementary cells of the barrier between cavities. Note that in each case input and output media have an impedance $Z_1$. In all the examples given in this figure, $p=3$ and $q=2$.}
\end{figure}

By inserting defects or impurities in the periodic structure as shown in Fig. \ref{Fig3}, it is possible to create localized modes whose resonant frequencies appear within the photonic band-gap\cite{Pradhan99,Luna_Acosta2008}. The defects have a length $\ell_D$ and an impedance $Z_D$. 
In this work we focus on structures with one, two or three identical defects.

\subsubsection{Photonic crystal cavities}

\begin{figure}
\includegraphics[width=8cm]{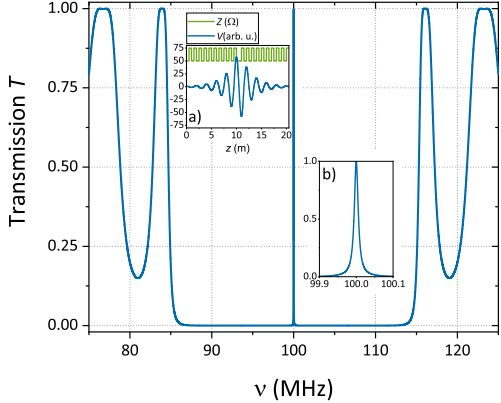}
\caption{\label{Fig4} Transmission of a one-dimensional photonic crystal cavity made with parameters given in the caption of Fig. \ref{Fig2} and $p=10$. Inset a) Impedance $Z(z)$ and voltage $V(z)$ spatial distributions given for $\nu=\nu_0$. Inset b) zoom of the transmission spectrum close to the resonance $\nu_0$.}
\end{figure}

The first structure is obtained by inserting a single defect\cite{Sanchez03}. The structure is described in Fig. \ref{Fig3}.a): it consists of  a one-dimensional photonic crystal with $N=2p$ cells where an extra layer of length $\ell_D=\ell_1$ and impedance $Z_D=Z_1$ is added after $p$ periods. This structure can also be seen as a single mode Fabry-Perot resonator made of two Bragg mirrors and a central layer of optical length $\frac{\lambda_0}{2}$. The matrix $\mathbf{M}_{1D}$ associated to this structure is given by:
\begin{equation}
\mathrm{\mathbf{M}}_{1D}=\mathrm{\mathbf{M}}^p\mathrm{\mathbf{Q}}_{1}\mathrm{\mathbf{M}}^p,
\end{equation}
and can be used to determine the spectral response of the defect-structure. Figure \ref{Fig4} shows the transmission spectrum of such a structure with $p=10$ corresponding to the periodic structure studied in Fig. \ref{Fig2} with a single defect.
Moreover, the inset a) of Fig. \ref{Fig4} gives the voltage (or field) $V(z)=V_+(z)+V_-(z)$ and impedance $Z(z)$ distributions inside the structure at $\nu=\nu_0$. An inspection of these two plots demonstrates clearly that a mode localized in the defect appears in the middle of the photonic bandgap at the frequency $\nu_0=100~\mathrm{MHz}$. The inset b) of Fig. \ref{Fig4} is a zoom of the transmission peak which displays a Lorentzian profile. This resonator can also be described in the coupled mode theory (CMT) framework\cite{Haus91,Dumeige08} by writing the evolution equation of the localized mode amplitude $a(t)$ shown in Fig. \ref{Fig3}.a):
\begin{equation}\label{CMT}
\frac{da}{dt}=\left(j\omega_0-\frac{1}{\tau}\right)a(t)+\sqrt{\frac{2}{\tau_e}}V_{in}(t),
\end{equation}
where $\omega_0=2\pi\nu_0$, $\frac{1}{\tau_e}$ is the coupling rate of the mode to the external media through the Bragg mirrors and $\tau$ is the mode amplitude lifetime. Since we have neglected the losses, we have $\frac{1}{\tau}=\frac{2}{\tau_e}$. The output signal can thus be written as $V_{t}(t)=\sqrt{\frac{2}{\tau_e}}a(t)$. In the stationary regime at angular frequency $\omega$, $a(t)=\tilde{a}e^{j\omega t}$, $V_{in}(t)=\tilde{V}_{in}e^{j\omega t}$ and $V_{t}(t)=\tilde{V}_{t}(\omega)e^{j\omega t}$. It is straightforward to solve Eq. (\ref{CMT}) and the transmission of the system then reads
\begin{equation}
T(\omega)=\left|\frac{\tilde{V}_{t}(\omega)}{\tilde{V}_{in}}\right|^2=\frac{\frac{1}{\tau^2}}{\left(\omega-\omega_0\right)^2+\frac{1}{\tau^2}}.
\end{equation}
The transmission resonance has thus a Lorentzian shape as shown in the inset b) of Fig. \ref{Fig4}, its width $\Delta\nu$ is related to the mode amplitude lifetime and the quality factor $Q$ of the cavity by 
\begin{equation}
Q=\frac{\omega_0\tau}{2}=\frac{\nu_0}{\Delta\nu}.
\end{equation}

\subsubsection{Two coupled photonic crystal cavities}\label{coupled2theo}

Figure \ref{Fig2}.b) shows a system composed of two coupled cavities. It consists of two identical defect separated by a barrier with $q$ cells\cite{Armitage98}.
\begin{figure}
\includegraphics[width=8cm]{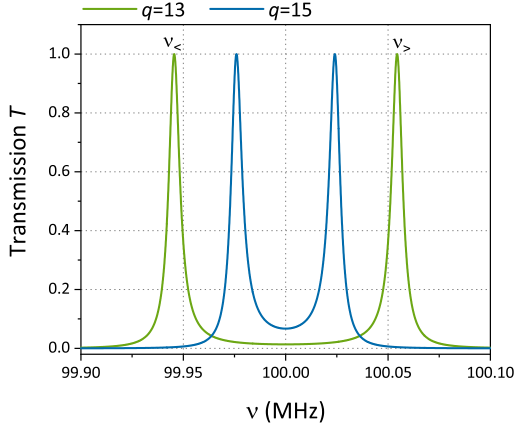}
\caption{\label{Fig5} Two coupled photonic crystal cavity transmission spectra for $q=13$ and $q=15$. In both cases the calculations have been carried out with $p=10$. $\nu_{>}$ and $\nu_{<}$ are respectively the high and low resonant frequencies of the two split modes.}
\end{figure}
The input and output coupling is obtained by tunneling through two barriers with $p$ periods. The matrix of the whole system reads
\begin{equation}
\mathrm{\mathbf{M}}_{2D}=\mathrm{\mathbf{M}}^p\mathrm{\mathbf{Q}}_{1}\mathrm{\mathbf{M}}^q\mathrm{\mathbf{Q}}_{1}\mathrm{\mathbf{M}}^p
\end{equation}
Calculations of the transmission are shown Fig. \ref{Fig5} for $q=13$, and $q=15$ in the case of 10-period input and output mirrors. The transmission spectra show that the initial resonance frequency $\nu_0$ is split into two high and low resonant frequencies. This can be understood by writing the evolution equations of the two resonant mode amplitudes\cite{Rasoloniaina15} $a_1$ and $a_2$ defined in \ref{Fig3}.b) via vector $\mathbf{a}=(a_1,a_2)^T$
\begin{equation}\label{CMT_vec}
\frac{d\mathbf{a}}{dt}=\mathbf{K}\mathbf{a}(t)+\sqrt{\frac{2}{\tau_e}}\mathbf{v}_{in}(t)
\end{equation}
where matrix $\mathbf{K}$ is given by
\begin{equation}
\mathbf{K} = 
\begin{pmatrix}
j\omega_0-\frac{1}{\tau} & j\gamma \\
j\gamma & j\omega_0-\frac{1}{\tau} \\
\end{pmatrix},
\end{equation}
and $\mathbf{v}_{in}(t)=(V_{in}(t),0)^T$. In this case $\tau=\tau_e$ and $\left|\gamma\right|$ is the coupling rate between the two cavities. The eigenvalues of $\mathbf{K}$ are $j\omega_{S,AS}=j(\omega_0\pm\gamma)-\frac{1}{\tau}$ and are associated to the symmetric and antisymmetric resonant modes of the whole system. As we consider a loss-less material, $\gamma$ is real; for an even value of $q$ we have $\gamma>0$ and the symmetric mode corresponds to the higher resonant frequency value thus $\nu_>=\frac{\mathrm{Re}(\omega_S)}{2\pi}$ and $\nu_<=\frac{\mathrm{Re}(\omega_{AS})}{2\pi}$, whereas for an odd value of $q$, $\gamma<0$ which leads to $\nu_>=\frac{\mathrm{Re}(\omega_{AS})}{2\pi}$ and $\nu_<=\frac{\mathrm{Re}(\omega_{S})}{2\pi}$. This control of the sign of the coupling coefficient is similar to what has been observed in two dimensional photonic crystal molecules\cite{Haddadi14}.
\begin{figure}
\includegraphics[width=8cm]{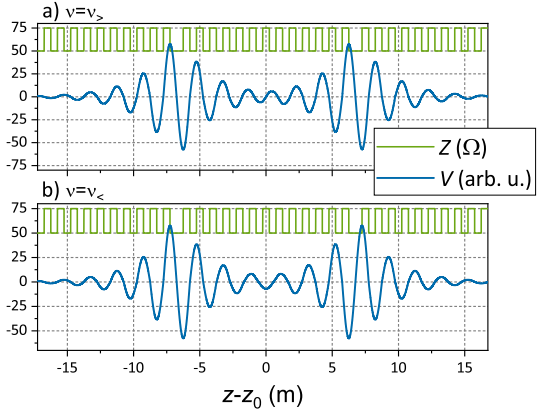}
\caption{\label{Fig6} Impedance $Z(z)$ and voltage $V(z)$ distributions for a double cavity system obtained for $p=10$ and $q=13$. a) $\nu=\nu_>$ antisymmetric mode; b) $\nu=\nu_<$ symmetric mode. $z_0$ is the position of the center of the structure.}
\end{figure}
Figure \ref{Fig6} gives the voltage distribution $V(z)$ for $p=10$ and $q=13$ for the two frequencies $\nu_>$ an $\nu_<$ described in Fig. \ref{Fig5}. Since $q$ is an odd number, the voltage profile is symmetric with respect to the center of the barrier for $\nu_<$ and antisymmetric for $\nu_>$ in both cases the two mode are localized at the defect locations. Moreover, by increasing $q$ the coupling between the two cavities is reduced which leads to a smaller frequency splitting as illustrated in Fig. \ref{Fig5}.

\subsubsection{Three coupled photonic crystal cavities}\label{coupled3theo}

\begin{figure}
\includegraphics[width=8cm]{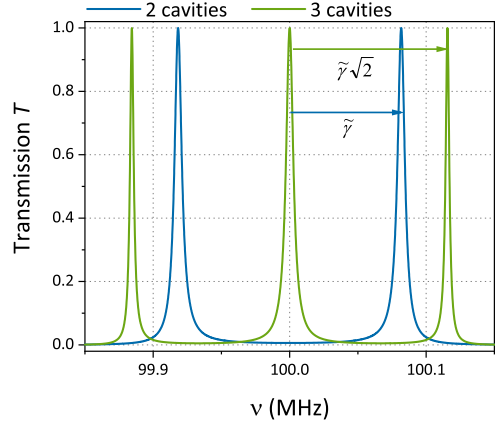}
\caption{\label{Fig7} Transmission spectra for $q=10$ and $p=10$ in the case of a double ($k=2$) and a triple ($k=3$) coupled cavity systems. $\widetilde{\gamma}=\frac{\gamma}{2\pi}$ is the frequency splitting in Hz.}
\end{figure}

This reasoning can be generalized to $k$ defects, in this case the transfer matrix $\mathbf{M}_{kD}$ is given by
\begin{equation}
\mathrm{\mathbf{M}}_{kD}=\mathrm{\mathbf{M}}^p\mathrm{\mathbf{Q}}_{1}\left(\mathrm{\mathbf{M}}^q\mathrm{\mathbf{Q}}_{1}\right)^{k-1}\mathrm{\mathbf{M}}^p
\end{equation}
In Fig. \ref{Fig3}.c) we have sketched a coupled cavity system with $k=3$ identical defects. The spectrum of such a system with $p=10$ and $q=10$ is shown in Fig. \ref{Fig7}. The initial mode is now split in three modes.
A more quantitative study of this splitting can be carried out using the CMT\cite{Armaroli15} using Eq. (\ref{CMT_vec}) where $\mathbf{a}=(a_1,a_2,a_3)^T$ (see Fig. \ref{Fig3}.c) and $\mathbf{v}_{in}(t)=(V_{in}(t),0,0)^T$. For $k=3$, matrix $\mathbf{K}$ is now given by
\begin{equation}
\mathbf{K} = 
\begin{pmatrix}
j\omega_0-\frac{1}{\tau} & j\gamma & 0\\
j\gamma & j\omega_0 & j\gamma \\
0 & j\gamma & j\omega_0-\frac{1}{\tau} \\
\end{pmatrix}.
\end{equation}
If the coupling rate is large enough ($\left|\gamma\right|\ll\frac{1}{\tau}$) the new resonant frequencies are the eigenvalues of $\mathbf{K}$ which read
\begin{equation}
\begin{pmatrix}
j\omega_0-\frac{1}{\tau} \\ j(\omega_0+\gamma\sqrt{2})-\frac{1}{2\tau} \\ j(\omega_0-\gamma\sqrt{2})-\frac{1}{2\tau}\\
\end{pmatrix}.
\end{equation}
Hence, the split resonances have a bandwidth which is half of that of the central resonance and the frequency splitting is increased by a $\sqrt{2}$ factor in comparison with the two cavity system. These effects are summarized Fig. \ref{Fig7} where we also represent the transmission of a double cavity obtained for $p=10$ and $q=10$. 

\section{\label{experience}Experiments}\label{experi}

\subsection{Experimental setup}

\begin{figure}
\includegraphics[width=8cm]{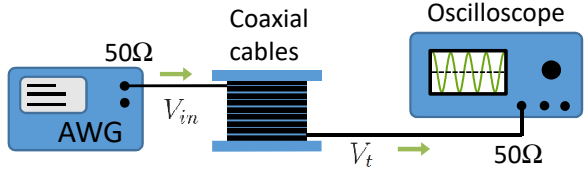}
\caption{\label{Fig8} Experimental setup. AWG: arbitrary waveform generator delivering an harmonic signal with an amplitude of $1~\mathrm{V}$. Coaxial cables: periodic or defect structures made of coaxial cables RG-58/U and RG-59/U.}
\end{figure}

Experiments have been carried out using two sets of 50~cm long RG-58/U and RG-59/U coaxial cables whose nominal impedances are respectively $50~\Omega$ and $75~\Omega$. Attenuation of cables are taken into account in the TMM by using the following values for the attenuation coefficients\cite{Radiall}
\begin{eqnarray}
\kappa_1(\nu) & = & 4.61\times 10^{-12}\nu+2.29\times10^{-6}\sqrt{\nu}\\
\kappa_2(\nu) & = & 4.61\times 10^{-12}\nu+1.46\times10^{-7}\sqrt{\nu},
\end{eqnarray}
where $\nu$ is given in Hz. For both cable sets the phase velocity is $v_{\phi 1}=v_{\phi 2}=0.66c$. The experimental setup is described Fig. \ref{Fig8}, the electromagnetic waves with frequency around $100~\mathrm{MHz}$ are produced by an arbitrary waveform generator SIGLENT SDG6022X and analyzed using an oscilloscope Tektronix DPO4104.
The transmission spectra are obtained by measuring the RMS values of the voltage at the input and at the output of the coaxial cable structures at several frequencies.

\subsection{Bragg mirror and photonic crystal cavity}\label{PCcav1Dexp}

\begin{figure}[h!]
\includegraphics[width=8cm]{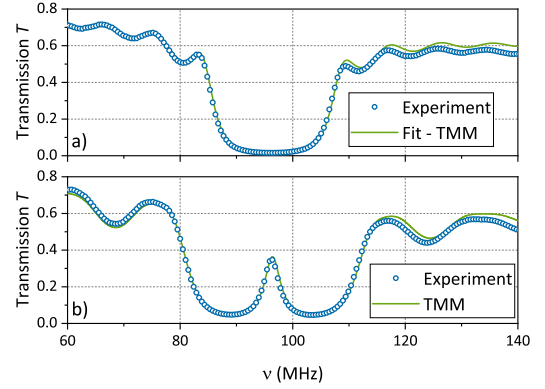}
\caption{\label{Fig9} a) Transmission of a periodic structure with $N=10$. b) Single defect structure ($k=1$) with $p=5$. Circles: measurements; full lines: calculations obtained by using the TMM.}
\end{figure}

The first experiment consisted in measuring the transmission of a periodic structure (see Fig. \ref{Fig1}.a) made of $N=10$ unit cells. The experimental results are given Fig. \ref{Fig9}.a). The actual length $\ell_1=\ell_2=51.9~\mathrm{cm}$ of the cables has been deduced by measuring the Bragg frequency $\nu_0=96.4~\mathrm{MHz}$ and by using Eq. (\ref{Bragg_cond}). The difference between the nominal and the actual cable lengths comes from the additional length due to the BNC adapters. The calculations shown in Fig. \ref{Fig9}.a) has been carried out using the TMM and varying $Z_1$ and $Z_2$. In the rest of this work we will use the impedance values $Z_1=54~\Omega$ and $Z_2=70.5~\Omega$ obtained from the best fit of the periodic structure transmission experimental data. The photonic crystal cavity or single defect resonant structure is obtained by adding an extra cable of impedance $Z_1$ (see Fig. \ref{Fig3}.a). The associated experimental transmission spectrum is given Fig. \ref{Fig9}.b).
At the center of the photonic bandgap ($\nu_0$), the transmission spectrum shows an attenuation-limited resonance with a spectral width $\Delta\nu=1.6~\mathrm{MHz}$ corresponding to a quality factor $Q=61.5$. The calculations (full line) which have been carried out using the length and impedance values obtained from the previous fit without adjusting any parameter are in good agreement with the experimental data.

\subsection{Two coupled photonic crystal cavities}

In this section we give the experimental results obtained for the system made of two cavities shown in Fig. \ref{Fig3}.b) and analyzed from a theoretical point of view at section \ref{coupled2theo}.
\begin{figure}[h!]
\includegraphics[width=9cm]{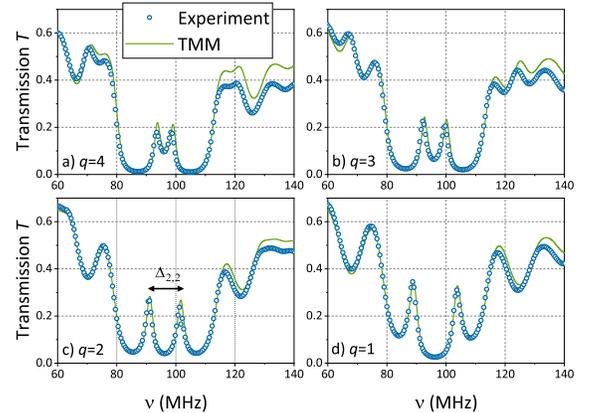}
\caption{\label{Fig10} Transmission spectra of two coupled cavity structures ($p=5$ and $k=2$) obtained for $q$ values ranging from 4 to 1. c) For $q=2$, the measured frequency splitting is $\Delta_{2,2}=10.8~\mathrm{MHz}$. Circles: measurements; full lines: calculations obtained by TMM.}
\end{figure}
The measurements have been done for several values of the number of unit cells separating the two cavities: $q\in\left\{1,4\right\}$. For $q=4$ (Fig. \ref{Fig10}.a), the cavity coupling is weak and thus the two split frequencies are not well separated. By decreasing $q$, the cavity coupling is increased leading to a stronger frequency splitting. In particular, for $q=2$ (Fig. \ref{Fig10}.c), we can measure a frequency splitting $\Delta_{2,2}=10.8~\mathrm{MHz}$ much larger than the single defect cavity resonance width. In four cases, the calculations plotted in full lines have been carried out using the TMM without adjusting the parameters found at section \ref{PCcav1Dexp}.

\subsection{Three coupled photonic crystal cavity}

The last experiment consisted in studying a three coupled cavity system ($k=3$) made of two unit cells in the barriers ($q=2$).
\begin{figure}
\includegraphics[width=8cm]{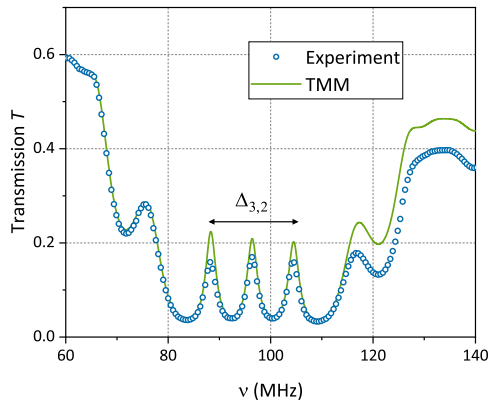}
\caption{\label{Fig11} Transmission spectrum of a three coupled cavity structure ($p=5$ and $k=2$) obtained with a barrier such as $q=2$. The experimental frequency splitting is $\Delta_{3,2}=16.1~\mathrm{MHz}$. Circles: measurements; full lines: calculations obtained by TMM.}
\end{figure}
The transmission spectrum is shown Fig. \ref{Fig11}. There is a good agreement between the numerical model (TMM) and the experiment. Nevertheless, since the structure is longer than in the previous experiments, the number of adapters increases and their attenuation should be taken into account to obtain a better agreement at higher frequencies. One can observe a splitting of the resonance frequency in three peaks as theoretically shown at section \ref{coupled3theo}. The measured value $\Delta_{3,2}=16.1~\mathrm{MHz}$ of this splitting is such as $\Delta_{3,2}/\Delta_{2,2}\approx1.49$ which is close to the theoretical value of $\sqrt{2}$ (see section \ref{coupled3theo}), the weak discrepancy comes from the fact that the attenuation has not been taken into account in the simple analytic theoretical model.

\section{conclusion}\label{conclusion}

We theoretically and experimentally analyzed the transmission properties of double and triple coupled photonic crystal cavities made of coaxial cables with impedance $50~\Omega$ and $75~\Omega$. We found good agreement between experimental data and simulated spectra obtained using the transfer matrix method. This model system is very versatile and easy to implement which could enable students to design and build themselves their own photonic structures during laboratory classes. This approach could be extended to chirped photonic structures or composite structures with more than two refractive indices by using cables with higher impedance. Moreover, the instruments required to carry out transmission experiments can be found in any teaching laboratory. Eventually, this approach can illustrate the coupling of resonators or oscillators which is a fundamental topic in physics teaching and goes beyond the electrodynamics or photonics frameworks\cite{Hansen96,Alzar02,Novotny10,Newman17}.

\begin{acknowledgments}
The authors acknowledge fruitful discussions with Ariel Levenson.
\end{acknowledgments}

%

\end{document}